# Interface-Confined Doubly Anisotropic Oxidation of 2-Dimensional MoS$_2$

Yejin Ryu[1], Wontaek Kim[1], Seonghyun Koo[1], Haneul Kang[1], Kenji Watanabe[2], Takashi Taniguchi[2] and Sunmin Ryu[1,3]*

[1]Department of Chemistry, Pohang University of Science and Technology (POSTECH), Pohang, Gyeongbuk 37673, Korea

[2]National Institute for Materials Science, 1-1 Namiki, Tsukuba 305-0044, Japan

[3]Division of Advanced Materials Science, Pohang University of Science and Technology (POSTECH), Pohang, Gyeongbuk 37673, Korea

**Abstract**

Despite their importance, chemical reactions confined in a low dimensional space are elusive and experimentally intractable. In this work, we report doubly anisotropic, in-plane and out-of-plane, oxidation reactions of 2-dimensional crystals, by resolving interface-confined thermal oxidation of a single and multi-layer MoS$_2$ supported on silica substrates from their conventional surface reaction. Using optical second-harmonic generation spectroscopy of artificially-stacked multilayers, we directly proved that crystallographically-oriented triangular oxides (TOs) were formed in the bottommost layer while triangular etch pits (TEs) were generated in the topmost layer, and that both structures were terminated with zigzag edges. The formation of the Mo oxide layer at the interface demonstrates that O$_2$ diffuses efficiently through the van der Waals (vdW) gap but not MoO$_3$, which would otherwise sublime. The fact that TOs are several times larger than TEs indicates that oxidation is greatly enhanced when MoS$_2$ is in direct contact with silica substrates, which suggests a catalytic effect. This study indicates that the vdW-bonded interfaces are essentially open to mass transport and can serve as a model system for investigating chemistry in low dimensional spaces.

**Introduction**

In atom-thin 2-dimensional crystals (2DXs), almost every atom resides on their surfaces. Therefore, their material properties are subject to various external perturbations. Owing to the sheet-like geometry and facile flexural deformation,[1] for example, 2DXs are often found conforming to underlying substrates[2] and mechanically strained.[3] Even a weak van der Waals (vdW) interaction with similar or dissimilar 2DXs or other materials modifies their electronic and vibrational structures.[4, 5] In addition, 2DX systems exchange energy efficiently with neighboring materials that have intimate contact. Interfacial thermal conductance between graphene and $SiO_2$ substrates is equivalent to that of carbon nanotubes dispersed in solution.[6] Transiently providing extra energy, charges are readily transferred across either direction between 2DXs and other chemical entities, as observed in time-resolved spectroscopy studies.[7, 8]

Owing to their high surface-to-volume ratio, surface chemical manipulation of one type of 2DX into another has been highly successful with methods such as oxidation,[9, 10] hydrogenation[11, 12] and halogenation.[13, 14] Despite the importance, mechanistic details of 2DX chemical reactions are far from being completely understood. Since many 2DXs are anisotropic in in-plane structure, they should exhibit edge-dependent reaction rates, which awaits direct observation. The presence of solid substrates that are required at least for mechanical stability creates another asymmetry between the top and bottommost layers. Possible reactions occurring at the hidden interface between 2DXs and substrates have not yet been under scrutiny and yet to be studied systematically. Since 2DXs are highly susceptible to external perturbations, the influence of substrates on their chemical behaviors is not a trivial problem. Mass transport through the interfacial space that is required for the reactions is not well understood.[15, 16] Moreover, the requirements of point defects as a reaction center need to be addressed.

In this work, for the first time, we report doubly anisotropic, in-plane and out-of-plane, reactions of 2DXs by resolving their substrate-mediated oxidation from their conventional surface oxidation. Crystallographically-directed in-plane anisotropic oxidation in 2D $MoS_2$ is initiated from preexisting defects and proceeds to form either triangular oxides or triangular etch pits. Using optical second-harmonic generation (SHG) spectroscopy combined with atomic force microscopy (AFM), we directly showed that both triangular objects are zigzag-edged and that quantified reaction rates varied ~3 fold depending on the crystallographic orientation of a given edge. We also showed that triangular oxides are formed in the bottommost layer of multilayer MoS$_2$ in direct contact with SiO$_2$ substrates at a 5-fold higher rate than triangular edge pits that are generated on the topmost layer, which constitutes the other anisotropy along the out-of-plane direction. The observed striking asymmetry demonstrates an active role of otherwise inert dielectric substrates and indicates that the vdW-bonded interface is essentially open to molecular diffusion at elevated temperatures.

**Results and Discussion**

Figure 1a shows the AFM height image of the oxidized single layer (1L) MoS$_2$ on SiO$_2$/Si substrates (see Methods). Unless noted otherwise, oxidation temperature ($T_{ox}$), time ($t_{ox}$), and molar fraction of O$_2$ ($x$(O$_2$)) were 320 °C, 2 hrs and 0.41, respectively. The 1L flake exhibited several triangular etch pits (TEs) of the same orientation and similar size as shown previously,[17-19] which can be attributed to anisotropic oxidative etching initiated at preexisting structural defects[17-19]: 2MoS$_2$ + 7O$_2$ → 2MoO$_3$ + 4SO$_2$. The lack of the oxidation products in the pits of TEs indicates that Mo oxides are volatile at the reaction conditions. Similar anisotropic etching was observed at the edges (denoted EE for edge-etching): the degree of etching marked by a pair of blue arrows varied significantly from one edge to another. Since the aligned TEs and varying rate of EE suggest in-plane anisotropy in the etching reaction as depicted in Fig. 1b, we employed SHG spectroscopy (Fig. 1c) to determine the preferred directions. Odd-number-layered 2H-MoS$_2$ belonging to the $D_{3h}^1$ space group[20] satisfies the non-centrosymmetry requirement for SHG but the even-numbered one ($D_{3d}^3$) does not. The polarized SHG signal, parallel to the polarization of 800 nm fundamental beam, is proportional to $cos^2 3\theta$,[21] where $\theta$ is the angle between the polarization vector and armchair direction ($\overrightarrow{AC}$) as shown in Fig. 1c. The SHG spectra obtained by rotating the sample (Fig. 1d) showed a periodic modulation which can be seen more clearly in the polar plot with the expected 6-fold symmetry (Fig. 1e). By comparing the SHG data with the AFM image, the orientation of $\overrightarrow{AC}$ was determined (Fig. S1) and denoted by double-headed arrows in Fig. 1a. Careful analysis of the AFM images (Fig. 1a) led us to the conclusion that the edges of the TEs are parallel to the zigzag (ZZ) directions. This agrees with the electron microscopy work on preferred edges of as-grown $MoS_2$.[22] In addition, ZZ edges terminated with Mo ($ZZ_{Mo}$) are known to be more stable than S-terminated ZZ ($ZZ_S$).[18] According to the consensus, the triangular objects observed in the current study were assumed to be $ZZ_{Mo}$-terminated, while there is some controversy.[17] Since the sample in Fig. 1a was crystallographically-resolved, the anisotropic EE at the peripheries of the crystal (Fig. 1a) allowed us to quantify, the reaction rates of the two distinct ZZ edges: $ZZ_{Mo}$ and $ZZ_S$ are, respectively, parallel and antiparallel to the edges of TEs as annotated in Fig. 1a & b. As shown in Fig. 1f, the oxidation occurred much faster at $ZZ_S$ than $ZZ_{Mo}$ for 1L at two different $O_2$ contents. As will be pointed out below, a similar contrast was observed for multilayer samples (2L ~ 4L; denoted as ML in Fig. 1f) and the average ratio of rate was 2.7 ± 0.6. The marked differential reactivity is a direct consequence of the in-plane anisotropy and will serve as an important number in testing possible reaction mechanisms.

In addition, we report another striking anisotropy along the stacking axis. The oxidized 2~4L crystals (Fig. 2a~d) showed small recessed triangles and large elevated triangles. Based on their phase information (Fig. S2), the former and latter were assigned as TEs and triangular oxides (TOs), respectively. The height profile lines showed that the depth of TEs was 0.80 ± 0.22 nm (3L in Fig. 2c) and 0.63 ± 0.28 nm (4L in Fig. 2d), which indicates that TEs are formed by oxidative etching of 1L, presumably the uppermost layer in multilayers, since the interlayer spacing of $2H-MoS_2$ is 0.62 nm. The TO areas were 0.36 ± 0.25 nm higher than the unreacted surface. The peripheries of 2L and 3L were elevated like TOs owing to edge oxide formation (EO) unlike the case of 1L that underwent EE. As will be shown later, the elevation is due to oxidation products, Mo oxides including α-MoO$_3$. Interestingly, the oxidation for TOs and EOs consumed 1L as shown by spatially-resolved Raman maps (Fig. S3). The rate of EO was also strongly dependent on the crystallographic orientation of a given edge (Fig. 1f). While both types of triangular structures were equilateral, TOs were approximately 5 times larger than TEs in multilayers. More surprisingly, both were aligned in the same (opposite) direction for odd (even)-numbered layers (Fig. 2c & d for 3L & 4L; see Fig. S4 for full AFM images). One might consider multiple preferred crystallographic orientations for TOs or TEs. However, there was only one 3-fold direction for either structure within a given crystal domain for all the samples. The clear even-odd alternation strongly suggests that both structures do not exist on the same uppermost layer. In a two-step, sequential oxidation of 3L, TEs and TOs were found to overlap with each other and grow independently of each other (Fig. S4). One plausible explanation is that TOs and TEs are respectively formed in the outermost layers as schematically depicted in Fig. 2e. In 2H-MoS$_2$ of odd (even)-numbered layers, the two surface layers are crystallographically oriented in a parallel (anti-parallel) manner, which agrees with the even-odd alternation. This scenario, however, requires that the MoS$_2$/SiO$_2$ interface is open to O$_2$ mass flow, required for the reaction, which is not well understood nor has been explored systematically.

To verify the aforementioned hypothesis, we fabricated an artificially-stacked (AS) quadrilayer (3L/1L) by the deterministic dry transfer[23] of 3L onto 1L. As indicated by $\overrightarrow{AC}$ in Fig. 3a, the crystallographic orientation of each flake was pre-determined using the SHG method and the twist angle between the top and bottom $\overrightarrow{AC}$ directions was set as ~30 degrees during the transfer step for the best angular resolution. Detailed procedures and optical micrographs of typical samples are given in Fig. S5. While the AS-3L/1L sample in Fig. 3b contained tiny trapped bubbles (Fig. 3c) with an interlayer spacing of 2.3 ± 0.2 nm, the interfacial coupling could be enhanced by transferring at an elevated temperature (Fig. S5). To introduce artificial defects as reaction centers within each region, mechanical nano-indentation was performed using specialized AFM tips (Fig. S6). TEs, TOs and an irregular etch pit were generated from the dents or natural defects in 1L (Fig. 3d), AS-3L/1L (Fig. 3f & g) and 3L (Fig. 3e) areas, respectively (see Fig. 3c for a large-scale image). The twist angle between $1L_{bottom}$ and $3L_{top}$ was 26.7 ± 0.5 degrees when measured after oxidation with SHG spectroscopy and their $\overrightarrow{AC}$ directions were denoted by the double-headed arrows in Fig. 3c & d. The edges of TEs in 1L were normal to its $\overrightarrow{AC}$ (Fig. 3d), confirming that they are ZZ-terminated. Notably, TOs found in AS-3L/1L (Fig. 3f & g) were of the same orientation as TEs in 1L and their edges were normal to the $\overrightarrow{AC}$ of 1L, not to that of 3L. This fact proves the hypothesis that TOs grow in the bottommost layer contacting the substrate unlike TEs that are formed in the topmost layer. The irregular shape in the 3L region is due to the randomly distributed bubbles and poor coupling with the substrate as will be explained below.

The formation of TOs in the bottommost layer indicates that oxygen molecules efficiently diffuse through the $MoS_2/SiO_2$ interface during oxidation but Mo oxides do not. To exclude the possibility of $SiO_2$ serving as an oxygen source,[24] control experiments were done in Ar gas, which did not appear to present oxidation (Fig. S7). Since the graphene/$SiO_2$ interface is tightly sealed against even He atoms,[15] we attribute the observed efficient interfacial mass transport to a thermally activated process. With increasing temperature, not only will the diffusion coefficient increase, but the vdW gap size at the $MoS_2$-substrate interface will also increase on average because of thermal expansion and fluctuate over time facilitating the mass flow. Oxygen molecules were found to diffuse through graphene/Ru at elevated temperatures and graphene/$SiO_2$ even at room temperature.[25, 26] Owing to the finite gap size, however, the interfacial diffusion will be greatly suppressed for larger molecules, which explains the formation of the interfacial Mo oxides. First, the almost complete EE indicates that oxidation products are volatile even at 320 °C. If $MoO_3$ is formed at the TO sites, its diffusion will be very slow since it is 4.5 times heavier than $O_2$. Moreover, the mass spectrometry study on thermal desorption of the $MoO_3$ powder showed that the most probable gaseous species is $(MoO_3)_3$, not the monomer.[27] More volatile suboxide clusters, $(MoO_{3-x})_n$, may also be formed from incomplete oxidation.[28] Thus, it is concluded that TOs are room-temperature condensates of gaseous Mo oxides mostly in the form of $\alpha$-MoO$_3$ (Fig. S8). High-resolution AFM images (Fig. 2c & d) also revealed that TOs are 2D arrays of nanoclusters and the clusters are radially aligned from the center of the triangles. The unique geometry may relate to inflation of the triangular MoS$_2$ membrane owing to the gaseous oxides and subsequent deflation induced by their condensation during the thermal cycle, but a detailed understanding requires further studies.

We also found that the size of vdW-gap indeed dictates branching between etching and oxide formation. To test this, an AS-2L sample of a good interlayer coupling was fabricated by stacking 1L on top of 1L/SiO$_2$ as shown Fig. S5b. Since the transfer was carried out at an elevated temperature to desorb possible adsorbate molecules on the sample surfaces, the AS region was virtually free of trapped bubbles and its interlayer spacing was 1.0 ± 0.2 nm, only slightly larger than that of 2H-MoS$_2$ (Fig. S5d). Despite the intimate coupling, the bottom 1L of AS-2L underwent EE and EO simultaneously, which stands in contrast with the predominant EO for the concealed bottom layer of 2H-2L and 3L (Fig. 2a). The concurrence of EE and EO indicates that the vdW gap space in AS-2L is somewhat efficient as a diffusion channel of Mo oxides, which is well corroborated by the measured interlayer spacing.

In addition, the MoS$_2$-SiO$_2$ interface plays a crucial role. As shown by the faster growth of TOs than TEs (Fig. 2), direct contact with SiO$_2$ substrates accelerates oxidation. Substrates may affect the chemical reactions of 2DXs by inducing structural deformation[10] or charge puddles.[29, 30] To unravel the role of substrates, MoS$_2$ samples supported on other substrates were prepared by the dry transfer method. When supported on crystalline hexagonal BN (hBN) and graphene, TOs were not generated (Fig. S9), which may be attributed to good coupling with the substrates (Fig. S5) that blocks interfacial diffusion of oxygen. Indeed, the average roughness of MoS$_2$ was significantly smaller on hBN than on SiO$_2$ substrates (Fig. S10). In addition, growth of TEs in 1L mechanically exfoliated on Si$_3$N$_4$/Si substrates was 7.5 ± 3.6 times slower than in 1L/SiO$_2$/Si (Fig. S11). Surprisingly, however, TEs of dry-transferred 1L/SiO$_2$/Si, which is in poor contact with SiO$_2$, were similar in size to those in multilayers, but 3.3 ± 1.8 times smaller than those of mechanically exfoliated 1L (Fig. S11). These results suggest a catalytic role of $SiO_2$ substrates and also indicate that an intimate contact with $SiO_2$ is necessary for the enhancement. While substrate-mediated charge doping and foreign species such as water trapped between $MoS_2$ and substrates may be responsible, detailed mechanism for the enhancement deserves further investigation.

Our work sheds some light on the nature of the defects that initiate growth of TEs and TOs. Although oxidations of 2DXs have been suggested to grow at preexisting defects, requirements as a seed are not known. 2D $MoS_2$ of various sources contain many different types of defects and simplest atomic defects such as vacancies and antisites are very much abundant ($0.01 \sim 0.1$ $nm^{-2}$), and more complex ones are less abundant.[31] However, TEs and TOs are very rare, being several orders of magnitude smaller in number density (approximately 1 $\mu m^{-2}$; Fig. S12 for statistics). Such scarcity would defy easy detection using probes with atomic resolution like transmission electron microscopy and scanning tunneling microscopy because of their limited screening speed. Thus our results imply that the defects responsible for TEs and TOs are ones with many atoms involved, not single or few-atom defects. Edges of flakes also serve as initiation defects for etching when exposed and oxide formation when buried (Fig. S13). Alternatively, tiny defects may coalesce to form defects that are larger enough to initiate oxidation through thermally activated diffusion, as suggested by a recent work.[32] The fact that TEs and TOs originating from the mechanical dents of approximately 10 nm in diameter were 30% larger (Fig. S6) implies that the initial stage of the reaction at the natural defects has higher activation energy possibly because of such an induction step.

The chemical reactions responsible for the TOs and EOs occur in quasi-2D space confined between $MoS_2$ and $SiO_2$ substrates and stand in contrast to conventional reactions in a few aspects. First, the transport of reactants and products is size-dependent. For conventional surface reactions, reactants and products exist either on surfaces or in the gas phase, and do not experience spatial restriction. For the confined reactions described in this work, one reactant ($MoS_2$) is partially isolated from the outer environment by highly flat but atomically rough $SiO_2$ substrates with typical roughness of 0.4 nm.[2, 33] Because of its small size, the other reactant ($O_2$) can be provided through the interfacial space between $MoS_2$ and $SiO_2$, which is not completely sealed against molecular diffusion. However, diffusional escape of the products ($Mo_nO_{3n}$) out of the confined space is very limited because of their large size, which led to triangular oxides (TOs). Diffusion of $SO_2$ will be more efficient. Secondly, the pressure within the confined space can be extremely high because of the 'van der Waals pressure'[34] induced by adhesion between $MoS_2$ and $SiO_2$. Recently, an internal pressure of 1 GPa was predicted for graphene-graphene space[34] and confirmed for graphene-$SiO_2$ space.[35] Lastly, $SiO_2$ substrates are physically contacting $MoS_2$ and may play a role in the reactions as suggested by results in Fig. S11. Further studies are required to reveal more details of confined reactions.

In summary, we reported asymmetric thermal oxidation reactions occurring at both surface layers of 2D $MoS_2$ crystals supported on $SiO_2$ substrates. TOs were formed at the bottommost layer while TEs were generated on the top. The crystallographic orientation of both triangles was determined for the first time using optical SHG spectroscopy. Formation of TOs demonstrated that oxygen inflow is highly efficient along the $MoS_2$/$SiO_2$ vdW interface, but outflow of gaseous Mo oxides was largely inhibited at an elevated temperature. The enhanced oxidation at the bottom layer was attributed to substrate-mediated charge doping or an unknown catalytic effect that awaits further investigation. This work presents intriguing aspects of chemical reactions confined at the $MoS_2$/$SiO_2$ interface and will contribute to understanding chemical processes in low dimensional materials and space.

**Methods**

**Preparation and treatments of samples:** Single and multi-layer samples of $MoS_2$, graphene and hBN were prepared by mechanically exfoliating their bulk crystals of high quality onto $SiO_2$/Si substrates.[36] The Si substrates were p-doped with a resistivity of 1 ~ 10 Ω·cm and covered with thermally grown 285-nm-thick $SiO_2$. Some samples were prepared on Si substrates with 100-nm-thick $Si_3N_4$ layer. Artificially stacked samples were prepared by a deterministic dry transfer method[23] as described in Fig. S5. Oxidation reactions were carried out in a quartz tube furnace (Lindberg, Blue M) at $T_{ox}$ = 320 °C for $t_{ox}$ = 2 hours unless otherwise noted. The molar fraction of oxygen gas, $x(O_2)$, in Ar carrier gas was varied in a range of 0.2 ~ 0.5, but 0.41 unless otherwise noted. The total pressure inside the tube was slightly above the atmospheric pressure and the flow rate of both gases were varied within 0.2 ~ 1.0 L/min to control the oxygen content.

**Raman spectroscopy:** Thin $MoS_2$ sheets of several μm across were identified under an optical microscope for fast screening and were characterized with Raman spectroscopy for their thickness and quality.[37] Raman spectra were obtained with a homebuilt micro-Raman spectrometer setup described elsewhere.[26] Briefly, a solid state laser beam operated at 514.300 nm was focused onto a sample with a spot size of ~1 μm using a microscope objective (40X, numerical aperture = 0.60). The average power on the sample surface was maintained below 0.2 mW to avoid any significant photo-induced effects. Back-scattered Raman signal was collected with the same objective and guided to a spectrometer combined with liquid nitrogen-cooled CCD detector. The overall spectral accuracy was better than 1 $cm^{-1}$.

**AFM measurements and nano-indentation:** The topographic details of samples were investigated using an atomic force microscope (Park Systems, XE-70). The height and phase images were obtained in a non-contact mode using Si tips with a nominal tip radius of 8 nm (MicroMasch, NSC-15). Nanometer-scale mechanical dents were generated on samples by exerting a force of 5 ~ 9 μN using an AFM tip (Budget Sensors, Tap300DLC) coated with 15-nm-thick diamond-like carbon. The nominal radius of the tip and force constant of the cantilever were <15 nm and 40 N/m, respectively. While reaction rates of EEs and EOs were defined as the displacement of edges per reaction time, those for TEs and TOs were defined as the radius of an inscribed circle divided by the reaction time.

**SHG spectroscopy:** Another microscope-based spectroscopy setup was used for the SHG measurements as described elsewhere.[38] Briefly, the output from a Ti:Sapphire laser (Coherent, Chameleon) operated at 800 nm was focused onto a 1.6 μm spot in FWHM with an objective lens (40X, numerical aperture = 0.60) as a fundamental beam. The duration and repetition rate of pulses were 140 fs and 80 MHz, respectively. The backscattered SHG signal centered at 400 nm was collected by the same objective lens and guided to a spectrograph (Andor, Shamrock 303i) equipped with a thermoelectrically cooled CCD (Andor, Newton). To vary the polarization angle of the incident fundamental beam with respect to the $MoS_2$ lattice, samples were rotated using a rotational mount with an angular accuracy better than 0.2°. Using an analyzing polarizer located in front of the spectrograph, either of the two perpendicular polarization components was detected.

## ASSOCIATED CONTENT

**Supporting Information**

Determination of crystallographic orientation, phase contrast & independent growth of TOs and TEs, intensity Raman map of TOs, independent growth of TOs and TEs, fabrication of artificially stacked layers, oxidation from mechanical dents, control reactions in Ar atmosphere, chemical nature of oxidation products, suppressed oxidation on non-silica substrates, roughness of samples and substrates, $MoS_2$ partially decoupled from substrates, number densities of TEs and TOs, exposed vs buried step edges. This material is available free of charge via the Internet at http://pubs.acs.org.

## AUTHOR INFORMATION

**Corresponding Author**

*E-mail: sunryu@postech.ac.kr

**Notes**

The authors declare no competing financial interest.

**ACKNOWLEDGMENTS**

This work was supported by the Center for Advanced Soft-Electronics funded by the Ministry of Science, ICT and Future Planning as Global Frontier Project (NRF-2014M3A6A5060934) and by the National Research Foundation of Korea (NRF-2016R1A2B3010390).

**FIGURE & CAPTIONS**

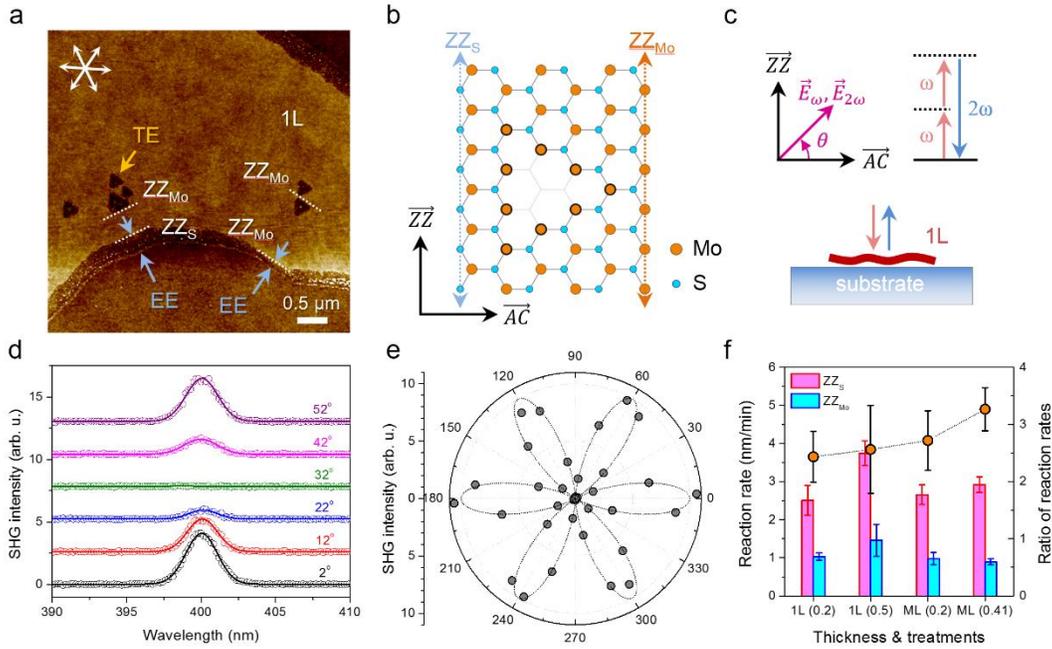

**Figure 1. Optical determination of in-plane anisotropy in oxidation of 1L MoS$_2$.** (a) AFM height image of the oxidized sample (x(O$_2$) = 0.20), revealing aligned triangular etch pits (TE) and edge etching (EE). Armchair directions ($\overrightarrow{AC}$) determined by SHG are denoted by the double-headed arrows. Zigzag edges terminated with Mo and S are denoted with ZZ$_{Mo}$ and ZZ$_S$, respectively. (b) MoS$_2$ lattice with ZZ$_S$ (left edge), ZZ$_{Mo}$ (right edge), AC edges (top and bottom) and ZZ$_{Mo}$-terminated TE. (c) Schematic view of SHG process and its measurements in a back-scattering geometry. The angle ($\theta$) between $\overrightarrow{AC}$ and SHG signal ($\vec{E}_{2\omega}$), which was parallel to input fundamental beam ($\vec{E}_{\omega}$), was varied. (d) SHG spectra obtained as a function of $\theta$. (e) Polar plot of SHG intensity. (f) (Left ordinate) Edge-resolved reaction rates of edge-etching (EE) of 1L and edge-oxidation (EO) of multilayers (ML: 2L ~ 4L): bar graphs. (Right ordinate) Ratios of reaction rates (ZZ$_S$/ZZ$_{Mo}$): orange circles. On average, reaction rate of ZZ$_S$ was 2.7 ± 0.6 times higher than that for ZZ$_{Mo}$. The numbers in parentheses of the abscissa are molar fraction of O$_2$.

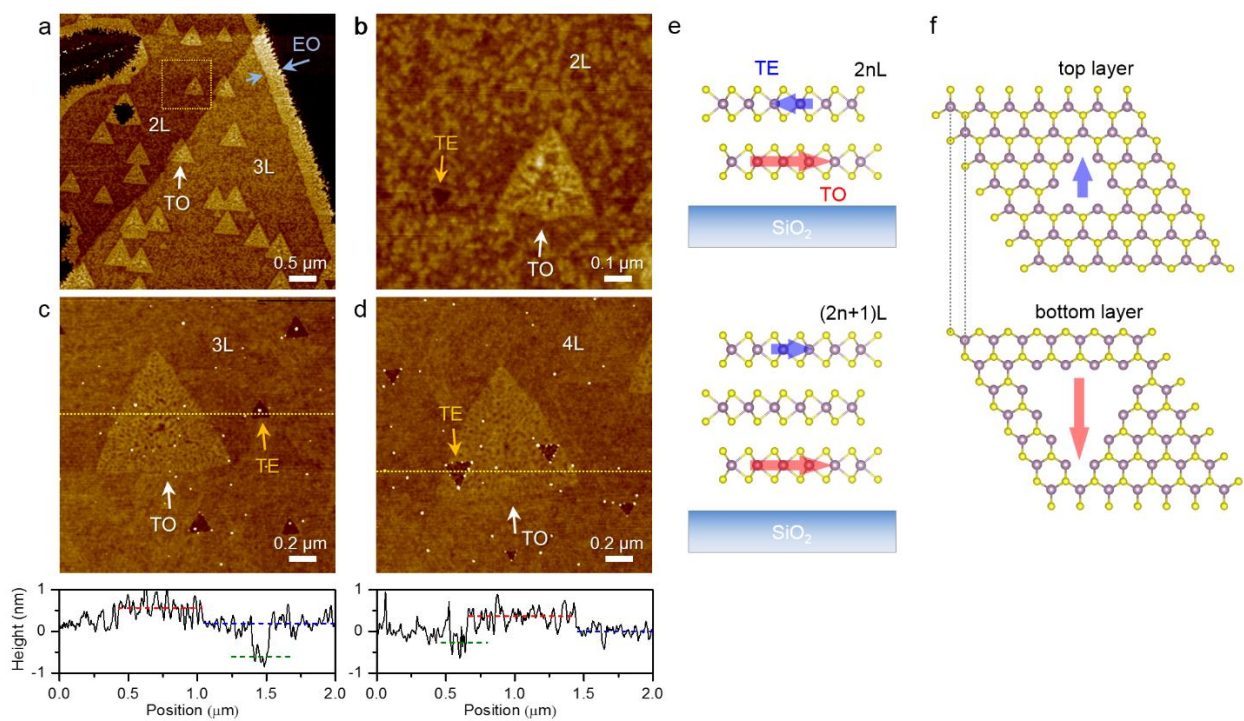

**Figure 2. Parallel and anti-parallel alignments of TE & TO.** (a) Height image of 2L & 3L with aligned triangular oxides (TOs) and EO. (b) Detailed image of an anti-parallel TE-TO pair in the square in (a). (c) Height image of a parallel TE-TO pair in 3L with height profile graph obtained along the yellow dotted line. (d) Height image of an anti-parallel TE-TO pair in 4L with height profile graph obtained along the yellow dotted line. Average height of unreacted, TO and TE areas (denoted by blue, red and green dashed lines in the profile graphs, respectively): $0.19 \pm 0.16$ nm, $0.56 \pm 0.20$ nm and $-0.61 \pm 0.15$ nm (3L); $0.00 \pm 0.12$ nm, $0.36 \pm 0.20$ nm and $-0.27 \pm 0.20$ nm (4L). $t_{ox} = 6$ hrs & $x(O_2) = 0.20$ for (a & b); $t_{ox} = 4$ hrs (c & d). See Fig. S4 for more TEs and TOs of 3L & 4L. (e) Schematic diagrams for the even-odd alternation of the orientations of TEs (denoted by blue arrow) and TOs (denoted by red arrow): even-numbered (2nL) layer (top), odd-numbered ((2n+1)L) layer (bottom). (f) The in-plane edge structures of TE and TO formed in 2L: TE in the top layer (top) and TO in the bottom layer (bottom). The inner triangular space of TO is filled with Mo oxides as described in the text. The vertical dotted lines guide the 2H-stacking in 2L.

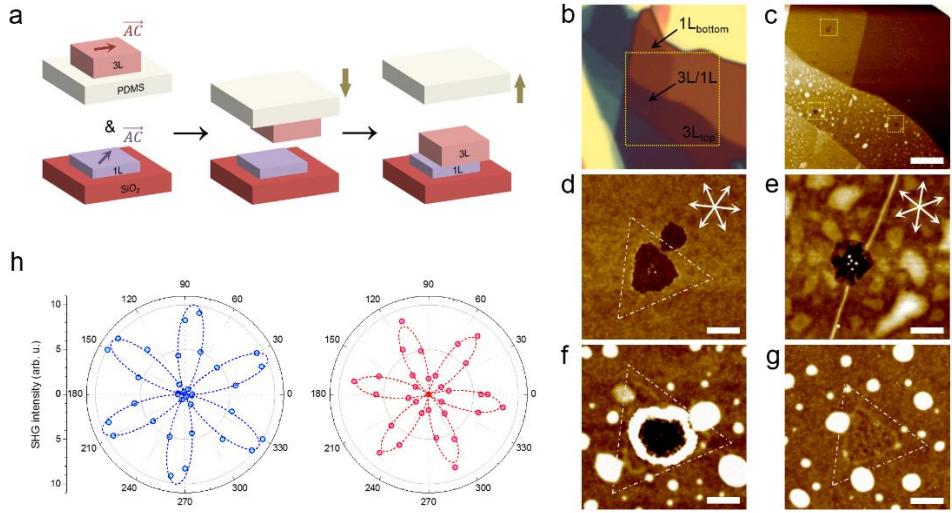

**Figure 3. SHG-determination of vertical anisotropy in oxidation of multilayer MoS$_2$.** (a) Simplified scheme of deterministic dry transfer to fabricate an artificially-stacked (AS) 3L/1L structure supported on SiO$_2$/Si substrate: (left) 3L and 1L were exfoliated on PDMS and SiO$_2$/Si substrates, respectively, followed by determination of $\vec{AC}$ with SHG spectroscopy; (middle) under an optical microscope, the 3L was brought into contact with 1L; (right) The PDMS substrate was detached gently leaving the AS-structure. (b) Optical micrograph of oxidized AS-3L/1L structure. (c) Height image obtained from the square in (b). The three squares denote where the indentation was made. (d-g) Height images obtained from the square areas in (c): TEs in 1L (d), etch pit in 3L (e), TOs in 3L/1L (f, g). (h) Polar plots of parallel SHG intensities from 1L (left) and 3L (right). TE (d), TO generated from a mechanical dent (f) and TO from a natural defect (g) were guided by dash-dotted triangles. Their edges were normal to the $\vec{AC}$ of 1L$_{bottom}$, not 3L$_{top}$ (double-headed arrows in d & e). Scale bars: (c) 2 μm, (d-g) 0.2 μm.

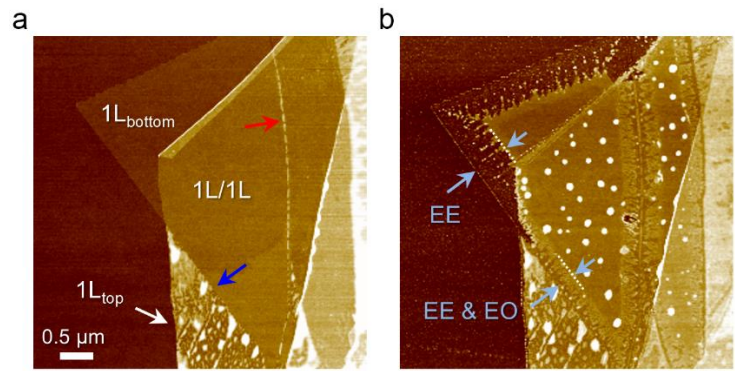

**Figure 4. vdW gap-controlled oxidation rate & diffusion of products.** a) Height image of pristine AS-1L/1L. The blue and red arrows denote an edge and crack line covered by $1L_{top}$, respectively. b) Image obtained after oxidation. While EE was predominant for uncovered $1L_{bottom}$, EE and EO occurred together for covered $1L_{bottom}$.

**Graphic TOC**

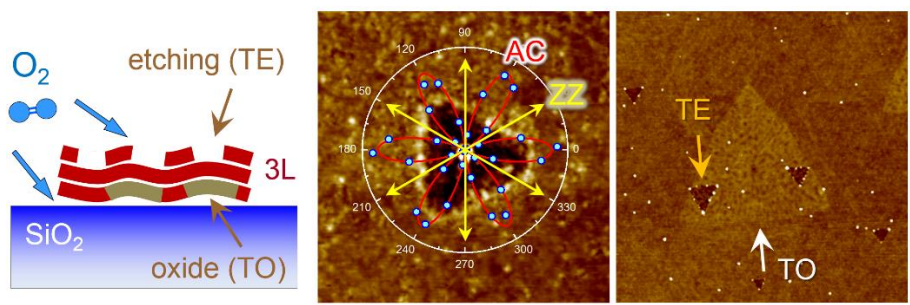